\documentclass[preprint,aps,showpacs,floats]{revtex4}
\usepackage{graphicx}

\def\lsim{\raise0.3ex\hbox{$\;<$\kern-0.75em\raise-1.1ex
\hbox{$\sim\;$}}}
\def\gsim{\raise0.3ex\hbox{$\;>$\kern-0.75em\raise-1.1ex
\hbox{$\sim\;$}}}

\begin{document}
 
\overfullrule 0pt

\title{
\vglue -1.8cm{\small \hfill IF - 1576/2003\\
\vglue -0.2cm
\hfill IFT-P.033/2003}\\
Discriminating among Earth composition models using geo-antineutrinos  
}

\author{H. Nunokawa$^{1,2}$}\email{nunokawa@ift.unesp.br}
\author{W. J. C. Teves$^3$}\email{teves@fma.if.usp.br} 
\author{R. Zukanovich Funchal$^3$}\email{zukanov@if.usp.br} 

\affiliation{\\ \\
$^1$ Instituto de F{\'\i}sica Te{\'o}rica,Universidade Estadual Paulista,
    Rua Pamplona 145, 01405-900, S{\~a}o Paulo, Brazil \\
$^2$ Departamento de F{\'\i}sica, 
Pontif{\'\i}cia Universidade Cat{\'o}lica do Rio de Janeiro, \\
C. P. 38071, 22452-970, Rio de Janeiro, Brazil \\
$^3$ Instituto de F{\'\i}sica,  Universidade de S{\~a}o Paulo 
C.\ P.\ 66.318, 05315-970, S{\~a}o Paulo, Brazil}

\begin{abstract}
It has been estimated that the entire Earth generates heat corresponding  
to about 40 TW (equivalent to 10,000 nuclear power plants) 
which is considered to originate mainly from the radioactive decay  
of elements like U, Th and K, deposited in 
the crust and mantle of the Earth. 
Radioactivity of these elements produce not only heat but 
also antineutrinos (called geo-antineutrinos) 
which can be observed by terrestrial detectors. 
We investigate the possibility of discriminating among 
Earth composition models predicting different total radiogenic 
heat generation, by observing such geo-antineutrinos 
at Kamioka and Gran Sasso, assuming KamLAND 
and Borexino (type) detectors, respectively, at these places. 
By simulating the future geo-antineutrino data 
as well as reactor antineutrino background contributions, 
we try to establish to which extent we can discriminate among 
Earth composition models for given exposures 
(in units of kt$\cdot$ yr) at these two sites on our planet. 
We use also information on neutrino mixing parameters coming 
from solar neutrino data as well as KamLAND reactor antineutrino data,  
in order to estimate the number of geo-antineutrino induced events. 
\end{abstract}
\maketitle
\thispagestyle{empty}

\section{Introduction}
\label{sec:intro}

There is much about the Earth's heat engine that is unknown. The main 
datum is the surface heat flow, as most of the Earth is hidden from view.
The total heat flux is presently estimated to be about 40 TW, 
which, however, suffers from uncertainties due to the size of its local 
variations and inaccessibility of much of Earth's surface.

We know that there are radioactive isotopes in the Earth that can produce 
heat through decay. Although the rate of heat generation by decay of unstable 
radioactive nuclides is tiny, if it is integrated over the entire volume of
the Earth, the total heat flux becomes huge.  Radiogenic heat, evidently, must 
be an important source of internal heat production.  However, questions as 
how much of the Earth's heat generation is from radiogenic origin, how much
from residual heat remaining from the formation of the Earth or from the 
release of gravitational energy as the Earth contracts, have not yet 
been answered in a satisfactory way.

In the context of modeling the thermal state and thermal history of the 
Earth it is important to know the specific heat production of the chief 
heat-producing radionuclides such as $^{40}$K, $^{238}$U, $^{232}$Th and 
$^{87}$Rb, encountered on surface layers and supposed to exist also in 
interior layers. The heat that drives mechanical motion in the mantle 
presumably comes mostly from radioactivity.  In the radioactive decay 
process, a portion of the mass of each decaying nuclide is converted 
to energy. Most of this is the kinetic energy of emitted $\alpha$
and $\beta$ particles or of $\gamma$ rays, which is fully absorbed by 
rocks within the Earth and converted to heat. 

For $\beta$ decays, however, 
part of the energy is carried away by emitted neutrinos and antineutrinos. 
Clearly, measuring the neutrino and antineutrino fluxes from the Earth can 
provide a unique way to access information on the internal structure 
and dynamics of our planet~\cite{Eder,Avilez,Krauss,Kobayashi,Chen,Ragh,
fiorentini1,fiorentini2,fiorentini3}.

This can be done, at least for
electron-antineutrinos ($\bar \nu_e$) coming from $^{238}$U 
and $^{232}$Th decays, 
the former producing 6 and the latter 4 $\bar \nu_e$ per decay chain 
within the energy reach of current and near future neutrino detectors.
Antineutrinos from these elements have higher energies than those that 
come from $^{40}$K and $^{87}$Rb, and are detectable at 
liquid scintillator detectors such as KamLAND~\cite{kamland} 
and/or Borexino~\cite{borexino}
through the inverse $\beta$-decay reaction, 
$p + \bar{\nu}_e   \to n + e^+$ where
the detection threshold energy is 
$E_{\bar \nu_e} = 1.7$ MeV.

Different Earth composition models predict different total 
amount of U and Th in the mantle, which lead to different 
total heat flux from radiogenic origin. 
Moreover, the fact that concentration of such elements is 
much larger (a factor of $\sim 20$ times) 
in the continental crust (with typical thickness 
$\sim 35$ km) than the oceanic one (with typical thickness 
$\sim 7$ km) and continents and ocean are not uniformly 
distributed over the Earth make the geo-antineutrino 
flux different from place to place. 

Quite recently, in Ref.~\cite{fiorentini1}, 
geo-antineutrino fluxes from various Earth composition models 
have been estimated (before the first KamLAND results were 
reported) and then in Ref.~\cite{fiorentini2}, the  
first KamLAND data, which contained $\simeq 9$ possible 
geo-antineutrino candidate events~\cite{kamland}, have been analyzed, 
and it was concluded that practically all Earth composition 
models are consistent with the current data. 
Moreover, the accumulated data so far at KamLAND is not enough 
to establish the presence of geo-antineutrinos
~\cite{kamland} and we must wait for future data. 

In this work, we will try to go beyond the above mentioned papers, 
by investigating to which extent liquid scintillator (type) detectors such
as KamLAND and/or Borexino can be used to help in 
discriminating among different geophysical models of heat production 
by measuring antineutrinos in the energy range 
$1.7 <E_{\bar \nu_e}/\mbox{MeV}<3.4$ 
produced inside the Earth. 
We show how much one can improve the quantitative understanding of 
the radiogenic contribution to terrestrial heat in about a decade of 
operation.

The organization of this paper is as follows.
In Sec.~\ref{sec:radio}, we briefly describe a few models for the
Earth radioactive element composition, 
in Sec.~\ref{anageo}, we present our analysis method and  
in Sec.~\ref{sec:resultados}, we describe our results. 
Finally, Sec.~\ref{sec:conclusion} is devoted to 
discussion and conclusion.

\section{Earth As a Antineutrino Source}
\label{sec:radio}

Antineutrinos are produced inside the Earth in radioactive $\beta$ 
decays mainly of $^{40}$K, $^{238}$U, $^{232}$Th and $^{87}$ Rb. 
These elements are classified as lithophile elements in 
geophysics and considered to be accumulated more in the Earth's crust. 
The abundance of these isotopes, although of prime geophysical importance,
is only known at or near the surface of the Earth. 
Among these geo-antineutrinos, the one come from 
$^{238}$U and $^{232}$Th have higher energies. 
The maximal neutrino energy from the former and the latter 
are, respectively, $E^{\rm max}_{\bar{\nu}_e}$ = 3.26 MeV 
and $E^{\rm max}_{\bar{\nu}_e}$ = 2.25 MeV, which are above
the threshold of existing anti-neutrino detector such 
as KamLAND~\cite{kamland} and Borexino~\cite{borexino}.

Here we consider, as our references, 
four different models which predict the distributions of 
$^{238}$U and $^{232}$Th in the Earth. 
We do not consider neutrinos coming from $^{40}$K and $^{87}$Rb because 
energies of these neutrinos are below the threshold 
of detectors we consider in this work. 
For all models, we have considered three regions, namely, continental 
crust, oceanic crust and mantle, with different average concentrations 
of U and Th. We ignore any contribution from the core. 
The distribution of radioactive elements are supposed to be uniform 
within each region.
Although the concentrations of these radioactive elements 
are considered to be much smaller in the mantle than 
in the crust, the total geo-antineutrino flux from the whole mantle 
can be comparable to that coming from the crust because of the
much larger volume.

Since the total mass of U in the crust is estimated to be
$M_{\text{c}}(\text{U})= 0.4 \times 10^{17}$ kg, and Si represents
about 15\% of the Earth mass, $M_\oplus=5.97\cdot 10^{24}\,$ kg, 
the other masses in the crust and in the mantle can be obtained for each
model from the mass ratios they provide. 
In this paper we will fix $M_{\text{c}}(\text{U})$ in the crust 
to the above estimated value, but we should point out that this 
number is known with 50\% uncertainty. The 
average concentration of U in the continental crust is estimated 
to be 1.7 ppm~\cite{wedepohl} whereas that in the oceanic one is 
estimated to be 0.1 ppm~\cite{taylor}. We will assume here these 
numbers as reference values but as we will discuss later it 
is important to know the local variations of $U$ concentration as well as
as the of the crustal thickness at each experimental site.

The models can be thus
classified by the amount of U they predict for the mantle
$M_{\text{m}}(\text{U})$.
In this work, we follow the classification of models considered 
in Ref.~\cite{fiorentini1,Ragh}, whose characteristic will be 
described below. 

\subsection{Chondritic Earth Model}
The Chondritic Earth model assumes for the Earth's gross composition
that of the oldest meteorites, the carbonaceous chondrites.  The mass
ratios for these meteorites~\cite{And} are:
$M(\text{Th})/M(\text{U})=3.8$, $M(\text{K})/M(\text{U})=7 \times
10^{4}$ and $M(\text{U})/M(\text{Si})=7.3 \times 10^{-8}$~\cite{Bro}.
Radiogenic production in the chondritic model easily accounts for 75\%
of the observed heat flow, about 30 TW. U and Th provide comparable
contributions, each a factor of two below that of K.
In this model, by taking into account that the total mass of 
the mantle is 4.1 $\times 10^{24}$ kg (68~\% of the total Earth mass), 
concentration of U in the mantle is about 0.006 ppm. 
\subsection{Bulk Silicate Earth (BSE) Model}
\label{bse}
The Bulk Silicate Earth (BSE) model provides
a description of  geological evidence  coherent with  geochemical
information. It describes the primordial mantle, prior to crust 
 separation. The mass ratios here are:
$M(\text{Th})/M(\text{U})=3.8$, $M(\text{K})/M(\text{U})= 10^{4}$ 
and $M(\text{U})/M(\text{Si})=9.4 \times 10^{-8}$.
In this BSE model the present radiogenic production, mainly from U 
 and Th, accounts for about one half  of the total heat flow, 20 TW.
 The antineutrino luminosities  from U and Th are rescaled 
 by a factor 1.3  whereas K, although reduced by a factor of 5, 
 is still the principal antineutrino source.
Concentration of U in the mantle is 0.01 ppm~\cite{Mc}. 

\subsection{Fully Radiogenic (FR I) Model}
\label{fully}
 One can conceive a model where heat production is
 fully radiogenic, with ${\rm K}/{\rm U}$ fixed at the terrestrial 
 value and  ${\rm Th}/{\rm U}$ at the chondritic value, which seems 
 consistent with terrestrial observations~\cite{fiorentini1}. 
 All the abundances are rescaled so as to provide the full  
 $40 \,{\rm TW}$ heat flow. All particle production rates  are correspondingly
 rescaled by a factor of two with respect to the predictions of the BSE
 model.   The concentration of U in the mantle for this model is 
 about 0.03 ppm. 

\subsection{Modified Fully Radiogenic (FR II) Model}
\label{secfully}

This model is similar to the previous one, the abundances of U and Th
are also rescaled with respect to BSE but it assumes that 
as an extreme case, the total
heat flow of 40 TW are produced in the Earth only by U and Th,
completely ignoring K, as considered in Ref.~\cite{Ragh}.
This is an extreme, geo-antineutrino fluxes much larger than this limit 
would need serious alteration of source distribution.
The concentration of U in the mantle for this model is about 0.04 ppm. 

\section{Analysis Method}
\label{anageo}

Here we explain how we calculate the expected number of geo-antineutrino
events, for each one of the four models presented in the previous section,
 at a certain detector position for a given exposure.
To try to distinguish models we have used a $\chi^2$ function 
minimization which is explained at the end of this section.

\subsection{Calculation of the antineutrino fluxes from Th and U decay 
chains}
\label{fluxgeo}

We would like to calculate the flux of antineutrino produced 
in the Earth by the decay of a certain isotope that reach a detector
for all four geophysical composition 
models presented in the previous section.
Following Refs.~\cite{fiorentini1,fiorentini2}, 
differential flux of antineutrinos produced in the decay 
chain of radioactive isotope $X$ that will be measured at 
a detector position $\vec{R}$ on the Earth, 
can be expressed by the following integral
performed over the Earth volume $V_{\oplus}$, 
\begin{equation}
\label{eq:eqflux}
\displaystyle \frac{d \Phi_{\bar{\nu}_e}(X)}{d E_{\bar \nu_e}} = 
\int _{V_{\oplus}} d^3r \frac{ \rho(\vec{r}) } 
{4\pi|\vec{R}-\vec{r}|^2} \,                                  
\frac{C(X,\vec{r})n_X}{\tau_X m_X}\,P_{\bar \nu_e}(E_{\bar \nu_e},|\vec r-\vec R|) \times f_X(E_{\bar \nu_e}), 
\end{equation}
where $\rho (\vec r)$ is the matter density, $C(X,\vec r)$, $\tau_X$, 
$m_X$ and $n_X$ are, respectively, the concentration, life time, 
atomic mass and the number of antineutrinos emitted per decay chain
corresponding to element $X={\rm U},\,{\rm Th}$.  
$f_X(E_{\bar \nu_e})$ is the normalized spectral function for 
element $X$~\cite{spectrum}, 
$P_{\bar  \nu_e}(E_{\bar \nu_e}, |\vec r-\vec R|)$ is the $\bar \nu_e$ 
survival probability, which can be averaged out, as a good approximation, and
bring out from the integral the term:
 \begin{equation}
\langle P_{\bar \nu_e} \rangle 
\simeq 1-\frac{1}{2} \sin^22\theta_\odot \simeq 0.58,
\end{equation}
where $\theta_\odot$ is the mixing angle responsible 
for the solar neutrino problem, which is 
fixed to the current best fitted value, 
$\sin^2 2\theta_\odot = 0.83$, 
obtained by combining the solar neutrino and KamLAND data~\cite{NTZ03},
except in Sec.~\ref{subsec:solar_angle} where 
$\theta_\odot$ is treated as a free parameter to be fitted. 
We note that the Earth matter effect on geo-antineutrino 
oscillation is very small and therefore, 
can be safely neglected in our analysis. 

The integration in Eq.~(\ref{eq:eqflux}) can be  
approximately divided into three distinct contributions, 
from continental crust (cc),   
oceanic crust (oc) and mantle (m),  
assuming that the matter density $\rho (\vec r)$ and the 
concentration of $X$, $C(X,\vec r)$ are approximately constant
within these three regions and 
ignoring any contribution from the core, 
as assumed in Ref.~\cite{fiorentini1}, as
follows, 
\begin{equation}
\Phi_{\bar{\nu}_e}(X) 
 \approx 
\frac{\langle P_{\bar \nu_e} \rangle}{4 \pi R_\oplus^2} 
\left[ L_{\bar \nu_e}^{\rm cc}(X) I_{\rm cc} 
+  L_{\bar \nu_e}^{\rm oc}(X) I_{\rm oc} 
+  L_{\bar \nu_e}^{\rm m}(X) I_{\rm m} \right],
\end{equation}
where $R_\oplus \approx 6370$ km is the radius of the Earth, and  we 
have defined the neutrino luminosity, $L_{\bar \nu_e}^i$,  and 
a factor which depend on the crust thickness distribution 
over the Earth as well as the detector location, 
$I_i$ ($i=$ cc, co, m), as
\begin{eqnarray}
\label{eq:lumi}
L_{\bar \nu_e}^i(X) & = & 
\frac{n_X \bar{C_i}(X) \bar{\rho}_i}{\tau_X m_X}, \\
\label{eq:I}
I_i & =& 
\frac{R_\oplus^2}{V_i}
\int_{0}^{2\pi}\,{\rm d}{\phi}
\int_{0}^{\pi}\,{\rm d}{\theta}\,\sin{\theta}
\int_{R_\oplus-h(r,\theta,\phi)}^{R_\oplus}\,{\rm d}r 
\frac{{r^2}}{{\left | \vec{R}-\vec{r} \right|}^2}\,,
\end{eqnarray}
where $V_i$, $\bar C_i(X)$ and $\bar \rho_i$ are, respectively,  
the volume, the average concentration of U or Th 
and matter density of region $i$. 
We can also write the  observable luminosities 
 $L_{\bar \nu_e}^i(X)$, in units of $10^{24}$ particles 
 per second, for masses in units of $10^{17}\,$ kg, as
\begin{equation}
\label{lnu1}
L_{\bar \nu_e}^i({\rm U})=7.4 \, M_i({\rm U})\, , \quad 
L_{\bar \nu_e}^i({\rm Th})= 1.6 \, M_i({\rm Th}).
\end{equation}

\begin{figure}
\vglue -1.0cm
\centering
\includegraphics[width=9.0cm]{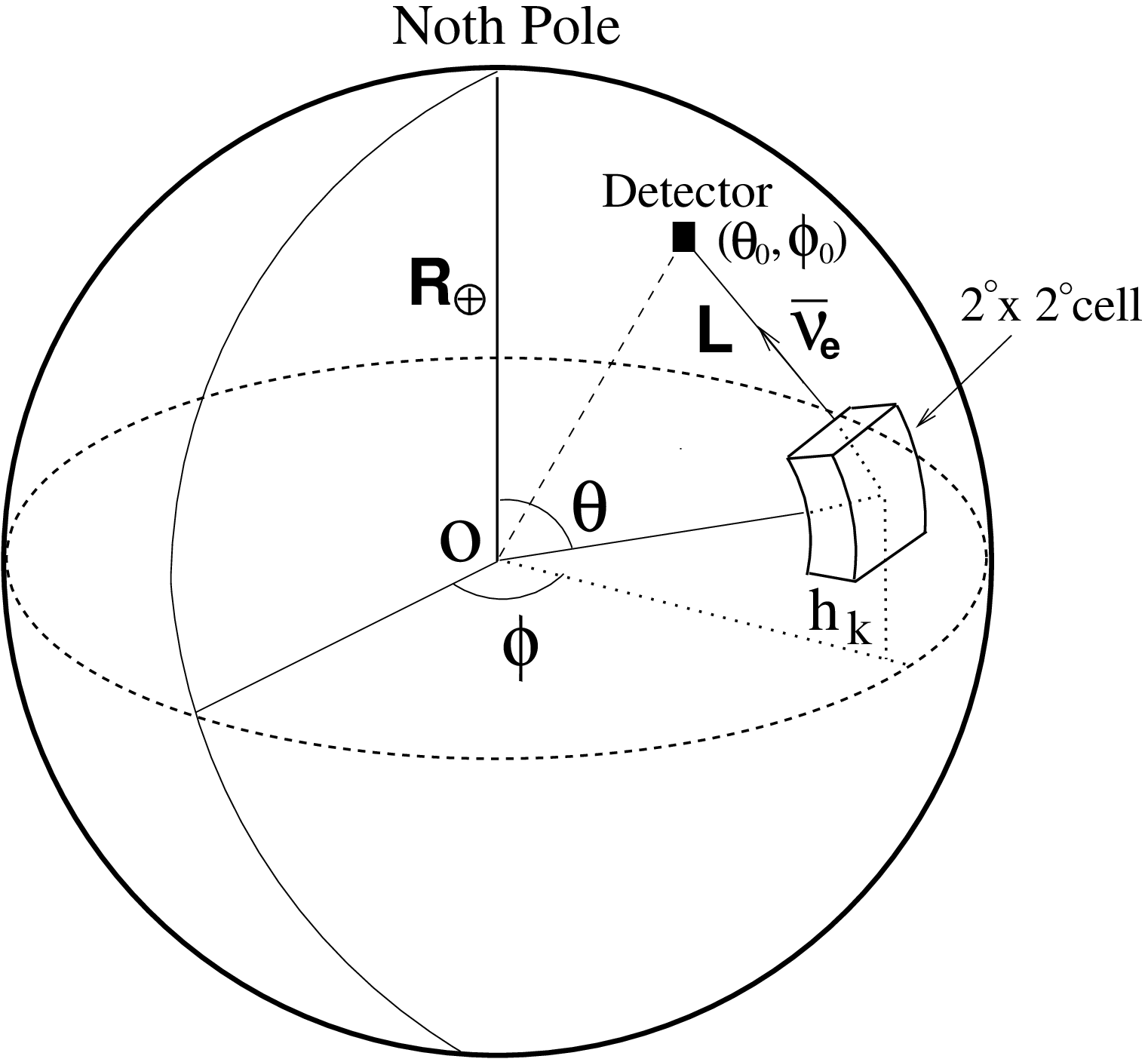}
\vglue -0.2cm
\caption{Schematic illustration of the {\it k}-th cell 
of the Earth crust in which its thickness is assumed 
to be constant. 
}
\label{fig:crust}
\end{figure}

\begin{figure}
\hglue -1.5cm
\centering
\includegraphics[width=18cm]{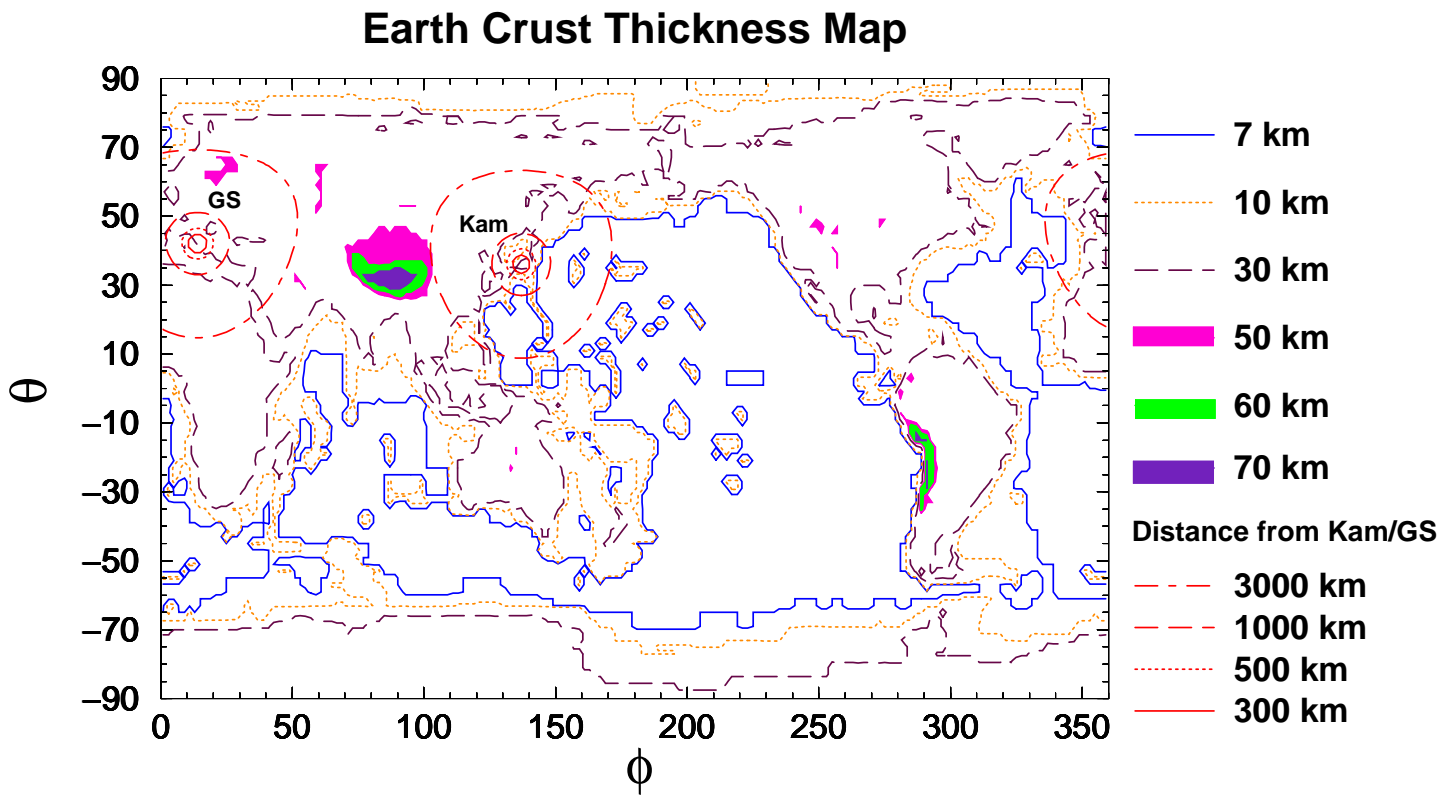}
\vglue -1.0cm
\caption{
Iso-contours of the Earth crust thickness based on 
the global $2^\circ \times 2^\circ$ crustal map \cite{ThicknessMap}
we adopted in this work.
}
\label{fig:map}
\end{figure}

In Eq.~(\ref{eq:I}), $h(r,\theta,\phi)$ indicates the
thickness of the shell (mantle or crust) at the 
position $(r,\theta,\phi)$. 
For the mantle we have simply used $h(r,\theta,\phi)=R_\oplus/2$, 
giving $I_{\rm m}=1.6$. 
For the crust calculation (typical thickness being $\sim$ 35 km for 
continental crust and $\sim$ 7 km for oceanic one), we have used 
the Earth Crust $2^\circ \times 2^\circ$ Thickness map \cite{ThicknessMap}, 
which was obtained based on seismology, to compute 
the $I_{\rm cc,oc}$ at a given detector position.
Earth's crust is divided into
16200 cells of $2^\circ \times 2^\circ$ where within 
each cell, the crust thickness is assumed to be constant. 
See Fig.~\ref{fig:crust} for a schematic illustration of
the cell. In Fig.~\ref{fig:map} we show some  
iso-contours of Earth crust thickness in the $\phi - \theta$ plane 
(corresponding to longitude and latitude) based on this crustal map 
model~\cite{ThicknessMap} which we will use in this work.

We observe that in our model assumption the predicted ratio 
of the geo-antineutrino fluxes from U and 
Th does not depend on details of the integration 
in Eq.~(\ref{eq:eqflux}) but is given by 
the following simple formula for any model we consider, 
\begin{equation}
\frac{\Phi_{\bar{\nu}_e}(\text{U})}
{\Phi_{\bar{\nu}_e}(\text{Th})} = 
\frac{M(\text{U})\ n_{\text{U}}\ \tau^{-1}_{\text{U}}}
{M(\text{Th})\ n_{\text{Th}}\ \tau^{-1}_{\text{Th}}}
\simeq 1.2\;.
\end{equation}
Note that we are assuming uniform U/Th distribution so that
the ratio of the total amount of Th and U,
$M(\text{Th})/M(\text{U})$, is considered to be 3.8 
in any of the input models under investigation.  

\begin{figure}
\vglue -1.8cm
\centering
\hglue -0.5cm
\includegraphics[width=25cm]{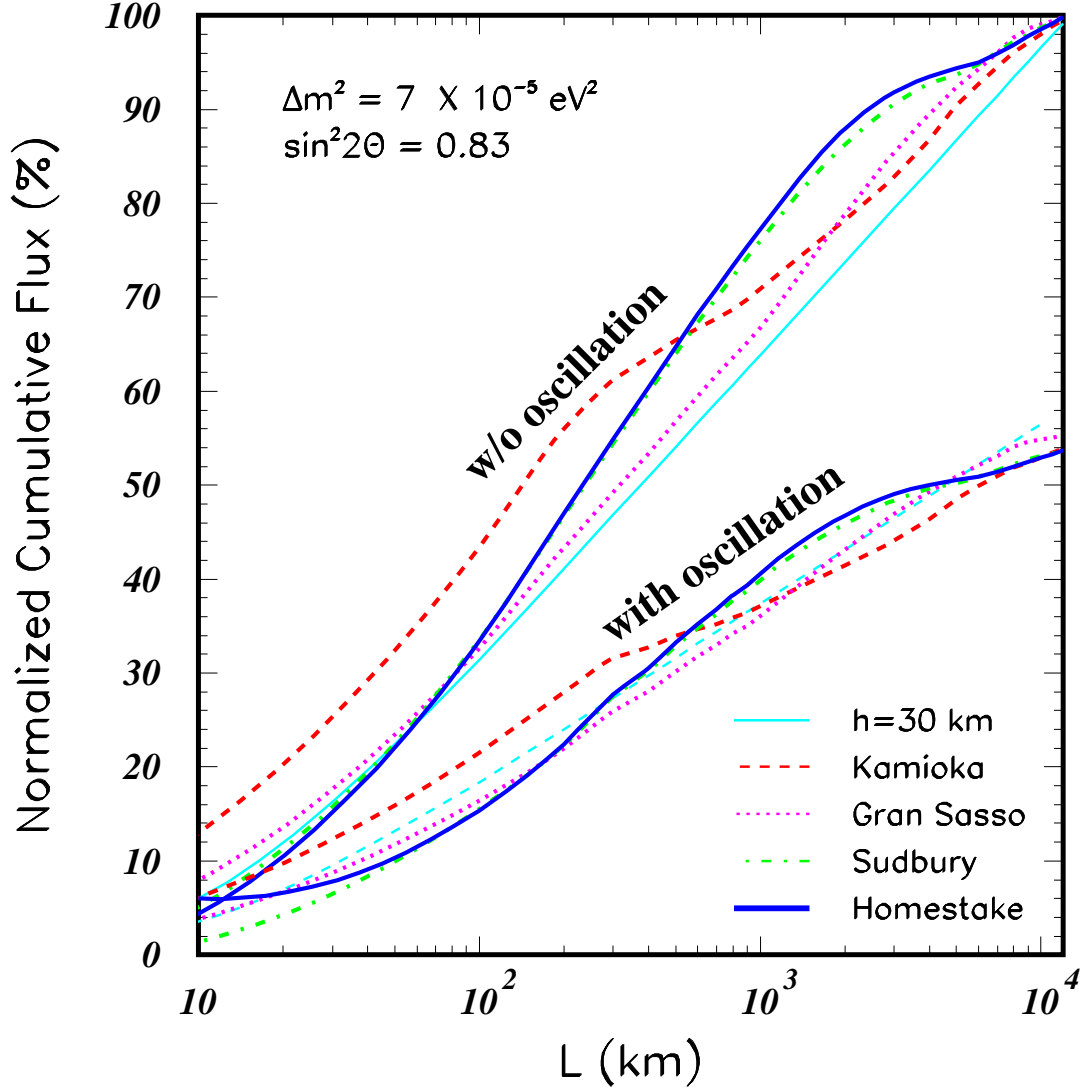}
\vglue -6.0cm
\caption{
  Normalized cumulative geo-antineutrino flux, with (lower five
  curves) and without (upper five curves) neutrino oscillation, coming
  from the continental as well as oceanic crust (without contributions
  from the Mantle) at various positions on the surface of the Earth as
  a function of the distance ($L$) from the source to the detector.
  These curves were computed using the information from the $2^\circ
  \times 2^\circ$ Earth Crust thickness map \cite{ThicknessMap}.  For
  the sake of comparison, we have also plotted the hypothetical case
  where the entire Earth crust has a uniform thickness of 30 km.  }
\label{fig:fluxo}
\end{figure}

We have thus the basic equations for determining radiogenic heat 
 production and neutrino flows from models of the Earth composition.
In order to have some feeling about the local variation of geo-antineutrino
fluxes due to the variable crustal thickness, 
in Fig.~\ref{fig:fluxo}, we present the 
normalized cumulative geo-antineutrino flux coming from 
the continental as well as oceanic crust (without contributions 
from the mantle) at Kamioka, Gran Sasso, Homestake and Sudbury, 
as a function of the distance ($L$) 
from the source to the detector, computed using 
the information from the 
$2^\circ \times 2^\circ$ Earth Crust thickness map \cite{ThicknessMap}.
We note that due to the fact that the abundance of U/Th 
is much larger in the continental crust than in 
the oceanic one, the geo-antineutrino flux from 
the former is dominant. 

Since the typical size of the $2^\circ \times 2^\circ$ 
crustal cell at these detector sites is of the 
order of 100 km, we can only reliably compute for distances equal or larger 
than this distance. 
In Fig.~\ref{fig:fluxo} we have extrapolated each curve 
down to 10 km, as a first approximation.
For the sake of comparison, we have also plotted the hypothetical case where 
the entire Earth crust has a uniform thickness of 30 km. 
Neutrino oscillations were not taken into account in the calculation 
of the upper five curves but were included in the lower five. These 
latter curves  have been normalized with respect to the no oscillation case. 

From this plot, as far as the geo-antineutrino flux coming from the 
Earth crust is concerned, we can see that about 30-40 \% of the total 
crustal antineutrino flux comes from a distance within 100 km 
and about 50-60 \% from a distance within 500 km from the detector.
We note that this flux is about 80\%,  65\%, 40\% and 33\%
of the total flux, respectively, for Chondritic, BSE, FR I and  
FR II models. 
This implies that it is very important to know rather well, with 
better than  $2^\circ \times 2^\circ$ resolution,
the variation of the crustal thickness near the 
detector as well as the local variation of the concentration of radioactive 
elements.

\subsection{Number of geo-antineutrino events}
\label{numbev}
The number of geo-antineutrino induced events from the decay
chain of element $X$ in the $i$-th energy bin, $N_i(X)$,  is:
\begin{equation}
\label{eqnuma}
N_i(X)= N_p \, t \, \int_i  dE_{\bar{\nu}_e}\, \epsilon(E_{\bar{\nu}_e})\, 
 \sigma(E_{\bar{\nu}_e})\,  \displaystyle \frac{d \Phi_{\bar \nu_e}(X)}{d E_{\bar{\nu}_e}} \,,
\end{equation}
where $N_p$ is the number of free protons in the fiducial volume of the 
detector, $t$ is the exposure time, $\epsilon$ is the detection efficiency,
which is assumed to be 100 \% for simplicity.
This integral, which is easily computed using the cross section given
in Ref.~\cite{cross}, is understood to be performed in a certain energy 
bin.

We divide the number of events in the positron prompt 
energy range 0.9-2.6 MeV into 4 bins with an interval of 0.42 MeV 
following Ref.~\cite{kamland}. 
Note that the prompt energy $E_{\text{prompt}}$ is 
related to neutrino energy as 
$E_{\text{prompt}} = E_{\bar{\nu}_e} - (m_n-m_p) + 2 m_e
=  E_{\bar{\nu}_e} - 0.78\ \text{MeV}$ where
$m_n, m_p$ and $m_e$ are respectively, mass of neutron, 
proton and electron. 
The number of events in $i-$th bin
is defined as the sum of events coming from 
U and Th,  $N_{\rm geo}(\bar E_i) \equiv N_i(\text{U}) + N_i(\text{Th})$.  
We note that the 1st two lower energy bins contains events coming 
from both U and Th induced geo-antineutrinos 
whereas the last two higher energy bins contain only events coming from Th.

\subsection{$\chi^2$ minimization}
\label{chi2}

We define the $\chi^2$ function for our geo-antineutrino analysis 
as follows, 
\begin{equation}
\chi^2_{\rm geo}
=  \sum_{i=1}^{17} 
\frac{\left [N^{\rm obs}(\bar{E}_i)-N^{\rm theo}_{\rm geo}(\bar{E}_i)-
N^{\rm theo}_{\rm reac}(\bar{E}_i)\right]^2}{N^{\rm obs}(\bar{E}_i)+\left (0.06\, N^{\rm obs}_{\rm reac}(\bar{E}_i)\right)^2},
\end{equation}
where $N^{\rm obs}(\bar{E}_i)$, $N^{\rm theo}_{\rm geo}(\bar{E}_i)$ 
and $N^{\rm theo}_{\rm reac}(\bar{E}_i)$ are respectively, 
the number of (simulated) observable events (geo + reactor antineutrinos), 
the number of theoretically expected events coming from geo-antineutrinos 
and from nuclear reactors in the neighborhood of the detector.
As mentioned before, geo-antineutrinos will only contribute to the 
first four bins. We have taken into account in our calculation the total 
statistical error as well as a 6\% systematic error only for reactor 
antineutrino events~\cite{kamland}. 
Since currently no detector has enough data in the energy region 
interesting for geo-antineutrino observations, we simulate  
$N^{\rm obs}(\bar{E}_i)$  according to the Earth composition models, 
taking into account the reactor antineutrino background for 
a given exposure and site. 
Here, for simplicity, we ignore the systematic error for geo-antineutrinos
as it is expected to be not so important compared to 
the statistical one for the exposure we consider in this work. 

The above $\chi^2_{\rm geo}$ will be minimized with respect to:
(i) $\Phi_{\bar \nu_e}(\rm Th)$ and $\Phi_{\bar \nu_e}(\rm U)$, the total
geo-antineutrino flux coming from Th and U at a given detector site;
(ii) $\Phi_{\bar \nu_e}(\rm U)$ assuming Th/U ratio fixed; 
(iii) $\sin^2 2\theta_{\odot}$ and $\Phi_{\bar \nu_e}(\rm U)$, 
assuming Th/U ratio fixed. 
In the first two cases the neutrino mixing parameters are
fixed to their best fitted values, in the latter only $\Delta
m^2_\odot$ is fixed (see the discussion in the following sections 
for further details).  
In addition, for some of our analyses, we have also added the $\chi^2$ 
function for solar neutrinos, $\chi_{\text{sol}}^2$, which was obtained in
Ref.~\cite{NTZ03}.

\section{Results}
\label{sec:resultados}
\subsection{Determination of Geo-antineutrino fluxes}

We first discuss the determination of geo-antineutrino fluxes
at Kamioka as well as Gran Sasso sites where the former (latter) 
has larger  (significantly smaller) reactor antineutrino 
background. 
In this work, we have assumed the reactor antineutrino flux at Gran Sasso 
5 times smaller than at Kamioka site, however, it can actually 
be even smaller~\cite{Ragh}.   
In this subsection, we assume that solar neutrino mixing
parameter will be determined with a good precision 
in the future. See, for instance, Ref.~\cite{sol_param} 
for a discussion on the perspectives of future determination of the 
solar neutrino mixing parameters.
Due to the effect of oscillation, the
observable flux suffers a reduction of $\langle P_{\bar\nu_e}\rangle =
1-\frac{1}{2}\sin^2 2\theta_\odot \simeq 0.58$, which implies that the
original flux must be understood as $\sim$ 1.7 times larger 
than we assumed at the detector site. 

Strictly speaking, experiments can only measure  
the total flux of geo-antineutrinos for a certain energy range,  
regardless of their origin (mantle, crust). They cannot directly 
access the amount of U and/or Th in the crust and 
mantle separately. Therefore, to be more conservative, 
we first try to consider the total geo-antineutrino fluxes from  
U and Th as free parameters to be fitted by the data 
for a given input model assumption. 
Since we still do not have enough data in the geo-antineutrino energy range 
($E_{\bar\nu_e}<3.4$ MeV), currently KamLAND has reported data only for about 
0.16 kt$\cdot$yr exposure~\cite{kamland}, 
we have simulated future data according to each one of the Earth composition 
models. We have used these simulated data points as an input in our analysis.
By performing a $\chi^2$ analysis we have investigated if the experiments can 
correctly reproduce the input data and distinguish among different models.

First we have estimated the required exposure, in units of 
kt$\cdot$yr to identify the presence of geo-antineutrino 
flux for an assumed model input, leaving the two components of 
the geo-antineutrino flux, coming from U and Th, completely free in our  
fit.
The results are shown in Table I.  
The numbers in the table are the detector exposure, 
at each site, required to rule out or accept each model at 3 $\sigma$ 
level.  
The actual fiducial volume of KamLAND (Borexino) is about 
0.4 (0.1) kt, which means one has to multiply by a factor 
2.5 (10) the number found in Table I in order 
to translate it to the actual detector exposure time.  
As naturally expected, the larger the input flux, the easier 
to identify the model, and therefore we need a smaller 
exposure to rule it out. 
If, for instance, a detector such as KamLAND, does not 
see geo-antineutrinos at the 3 $\sigma$ level after 0.46 kt$\cdot$yr 
exposure, it can rule out the FR II model, but at this point it cannot 
say anything about the other models. On the other hand, if 
KamLAND measures geo-antineutrinos at this exposure, then FR II 
has to be interpreted as the preferred model, and so forth.

\begin{table}
\caption[aaa]{
Required exposure in units of kt$\cdot$yr 
to identify geo-antineutrinos at the 3 $\sigma$ level, 
at Kamioka and Gran Sasso for each input 
model we consider in this work. 
}
\vglue 0.1cm
\begin{tabular}{|c|cccc|}
\hline
   &  Chondritic\ \    &  BSE  \ \ &  FR I \ \ &  
FR II \ \ \\
Site   &    &   &   &   \\
\hline
Kamioka      &  1.7  &  1.3  & 0.67   &  0.46 \\
Gran Sasso   & 0.89  &  0.71  & 0.38  &  0.26 \\
\hline
\end{tabular}
\label{Tabzero}
\vglue 0.5cm
\end{table}

\begin{figure}
\vglue -1.5cm
\centering\leavevmode
\hglue -2.3cm
\includegraphics[width=23cm]{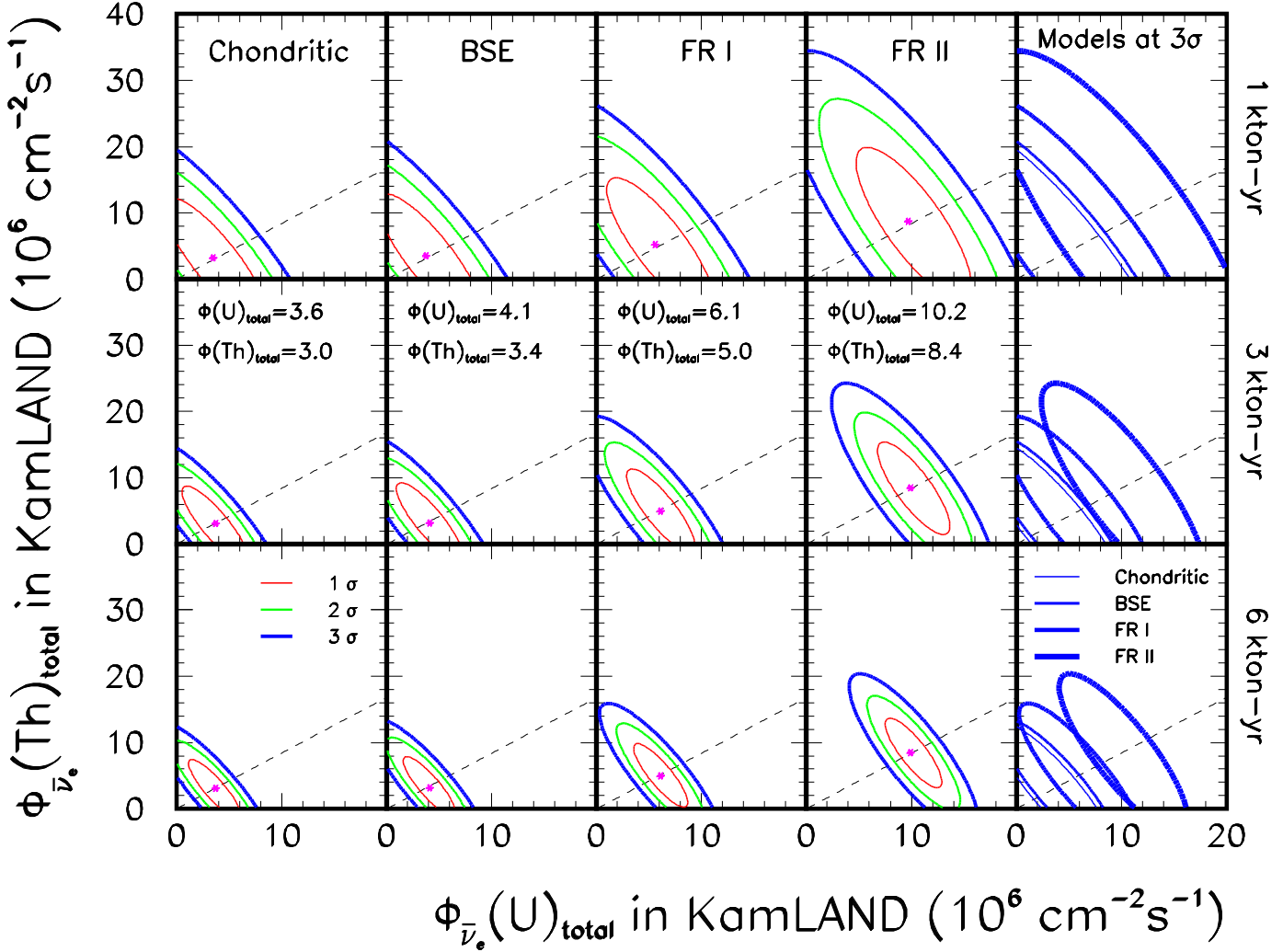}
\vglue -5.5cm
\caption{
Region allowed at 1, 2 and 3 $\sigma$ in the 
$\Phi_{\bar \nu_e}(\text{U})-\Phi_{\bar \nu_e}(\text{Th})$ plane, 
for 4 different models (in first 4 columns) 
for 1, 3 and 6 kt$\cdot$ yr of exposures 
(in different rows). In the 5th column, 
we have superimposed 3 $\sigma$ allowed regions
of 4 different models.  
In the second row, for each model, the 
input values for the geo-antineutrino 
fluxes computed according to Eq.~(\ref{eq:eqflux}) using 
the information from the Earth Crust 
$2^\circ \times 2^\circ$ Thickness map \cite{ThicknessMap}
are indicated. 
Both of $\Phi_{\bar \nu_e}(\text{U})$ and $\Phi_{\bar \nu_e}(\text{Th})$ fluxes are allowed to vary freely.  
Dashed lines indicate the expected U and Th induced 
antineurino fluxes if the Th/U mass ratio is assumed 
to be constant, $M(\text{Th})/M(\text{U})=3.8$.
The input points, which practically coincide with the best fit, 
are indicated by stars.  The input fluxes are estimated in the 
presence of neutrino oscillations with $\Delta m^2=7 \times 10^{-5}$ eV$^2$ 
and $\sin^2 2\theta = 0.83$.
}
\label{fig:phi_phi_kam}
\end{figure}

In Fig.~\ref{fig:phi_phi_kam}, we present in the $\Phi_{\bar
\nu_e}(\text{U})-\Phi_{\bar \nu_e}(\text{Th})$ plane, to which extent
we can determine geo-antineutrino fluxes in the presence of neutrino 
oscillations for given input model fluxes and
exposure (in units of kt$\cdot$yr) at the Kamioka site.  
In all cases we are able to correctly reproduce the input fluxes
as the best fit points (indicated by stars) in our fit, which, however,
 does not necessarily mean we can distinguish models with enough significance
(see the text below).
Unfortunately it seems to be extremely difficult to distinguish between 
BSE and Chondritic models, independently of the detector exposure we 
have considered here.
From this plot, we can conclude that for 1 kt$\cdot$yr of 
exposure, it is not possible to distinguish among the 
four models considered. 
However, after 3 kt$\cdot$yr of exposure, 
one start to have some sensitivity to distinguish models,
namely, Chondritic/BSE from FR II. 
After the maximal exposure considered, 6 kt$\cdot$yr of data, 
one can almost always distinguish FR II from any of the other 
models. It is worthwhile  to note that even for the maximal exposure 
and maximal flux (FR II) considered here, it is not possible to exclude 
a null Th induced geo-antineutrino flux, whereas 
it is sometimes possible to exclude a null U induced one.

\begin{figure}
\vglue -1.5cm
\centering\leavevmode
\hglue -2.3cm
\includegraphics[width=23cm]{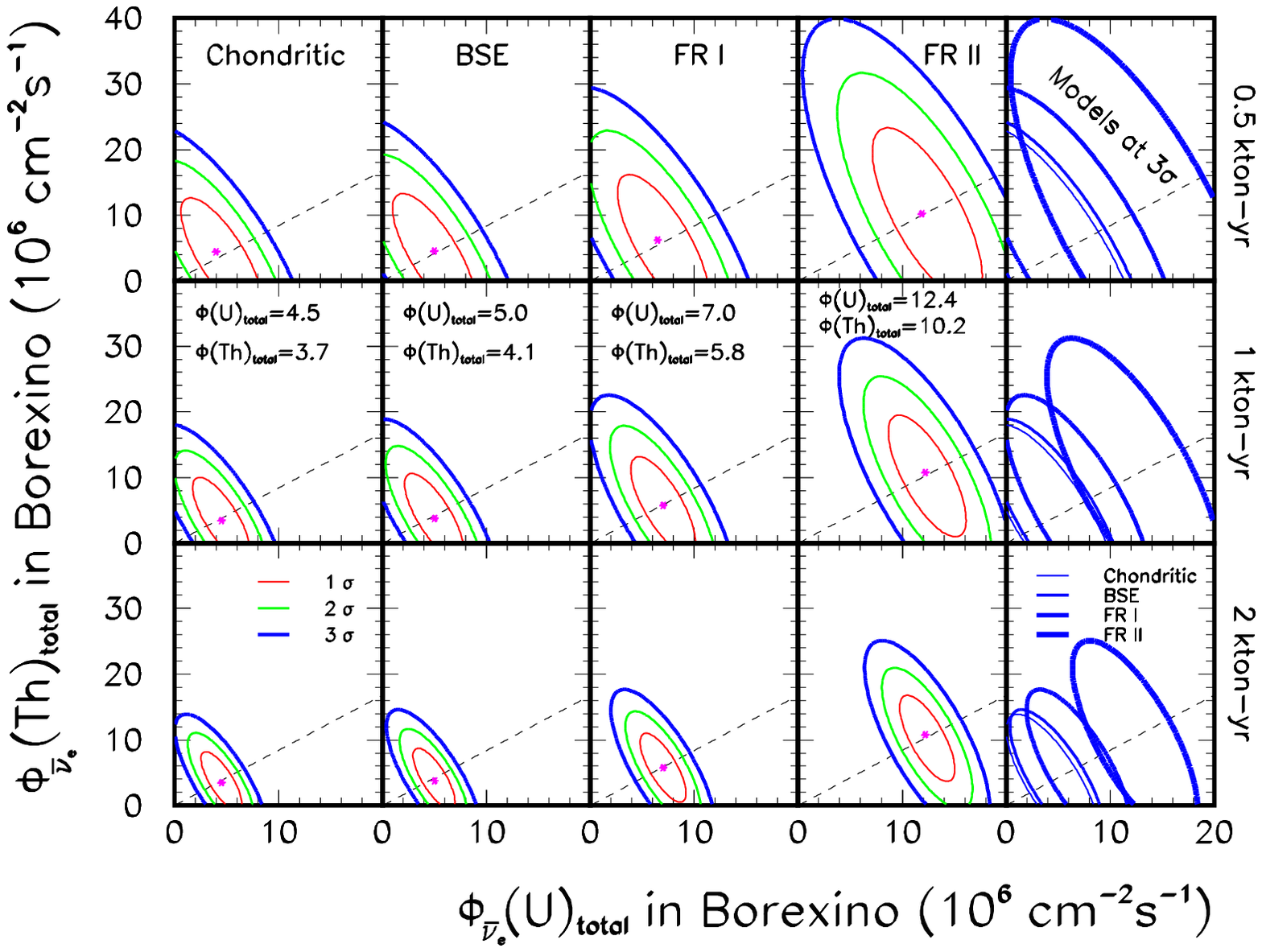}
\vglue -5.5cm
\caption{
Same as in Fig.~\ref{fig:phi_phi_kam} but for the Gran Sasso
site. 
}
\label{fig:phi_phi_gs}
\end{figure}

In Fig.~\ref{fig:phi_phi_gs}, we present the same plot
but for the Gran Sasso site, assuming a significantly less 
reactor neutrino background, 5 times smaller than 
at the Kamioka site.
In this case, one can achieve the same sensitivity 
for the flux determination with much less exposure compared
to the Kamioka site.
Roughly speaking, $x$ kt$\cdot$yr exposure at Gran Sasso 
correspond to $(2-3)x$ kt$\cdot$yr at Kamioka, which 
we can see by comparing 
Figs.~\ref{fig:phi_phi_kam} and ~\ref{fig:phi_phi_gs}.
The general behavior of the allowed regions are very similar to 
that for Kamioka site apart from this difference.

\begin{figure}
\centering\leavevmode
\hglue -2.3cm
\includegraphics[width=23cm]{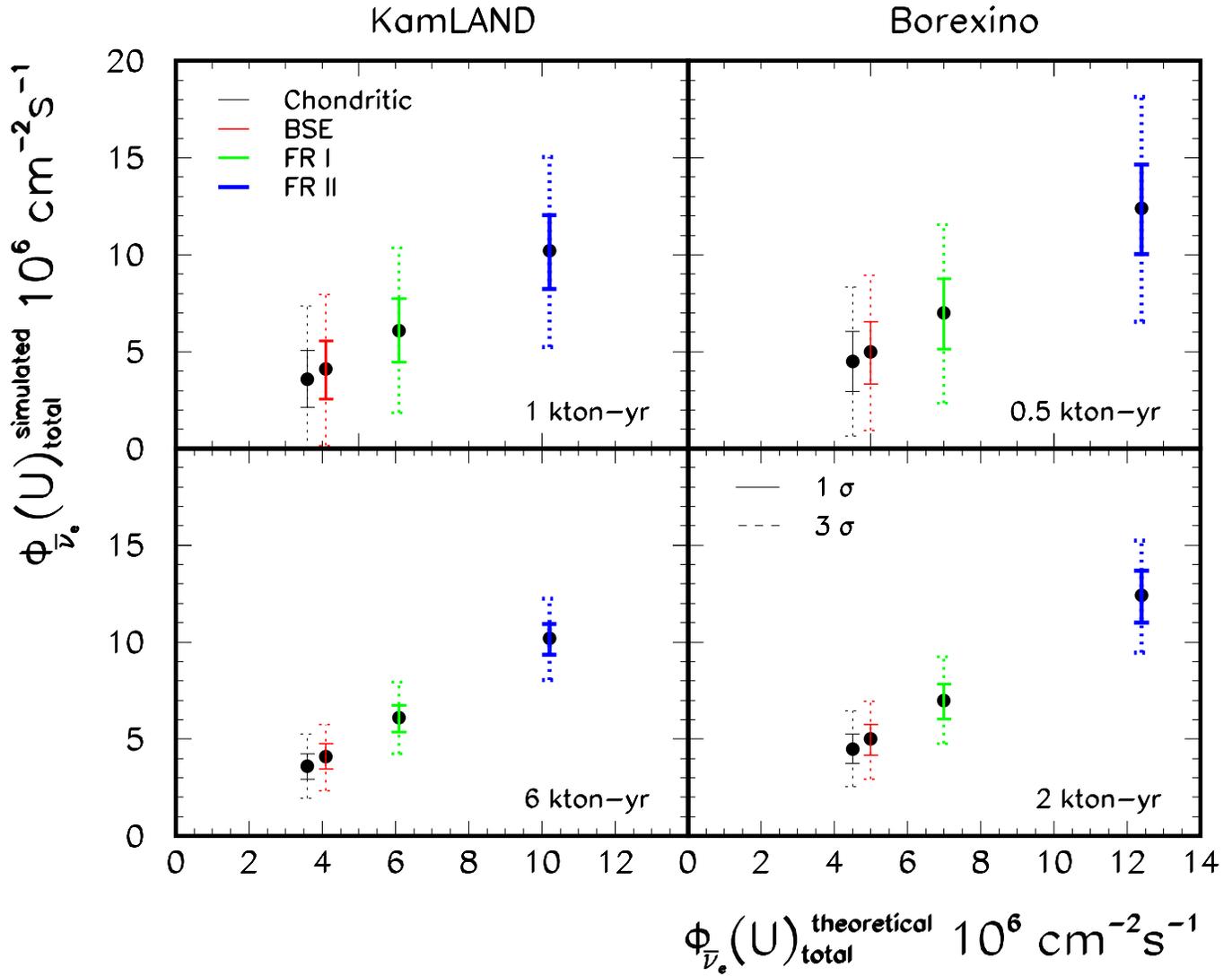}
\vglue -5.5cm
\caption{
The allowed ranges of geo-antineutrino flux from U are
plotted for given different input flux assuming 
$M(\text{Th})/M(\text{U})=3.8$ for KamLAND 
for 1 and 6 kt$\cdot$ yr of exposures 
and for Borexino for 
0.5 and 2 kt$\cdot$ yr exposures. 
}
\label{fig:modelos-fin}
\end{figure}

In Fig.~\ref{fig:modelos-fin}, we show with which precision
the total U induced antineutrino 
flux can be determined by the experiments, further
imposing that the ratio between the total U and Th mass in the Earth
is fixed to be constant, $M(\text{Th})/M(\text{U})=3.8$, in the
$\chi^2$ fit.  Note that this reasonable assumption, based on meteorite and  
Earth surface data,  is not very important in distinguishing models but 
allows for a better precision in flux determination.
In this case, we have only one free parameter,
$\Phi_{\bar\nu_e}(\rm U)$, to be fitted. 
In 6 kt$\cdot$yr at Kamioka, the total U flux can be determined within 
less than 10 \% uncertainty, 
independently of the model. 
The precision it can be determined at Gran Sasso, 
after 2 kt$\cdot$yr exposure, is a bit over 10 \%.

Let us now translate the flux uncertainties we have estimated 
into heat uncertainties, in order to compare our results 
with those of Ref.~\cite{fiorentini3}. As one can easily understand 
from Eq.~(8) of that paper, we can write the total heat, $H$, as 
a function of the geo-antineutrino flux as

\begin{equation} 
H = a \Phi_{\bar \nu_e} - b \;,
\end{equation}
where $a$ and $b$ are positive constants satisfying  $a \Phi_{\bar \nu_e}  \geq  b$. 
This implies that the relative heat uncertainty is always greater than the 
relative flux uncertainty, that is,

\begin{equation} 
\displaystyle \frac{\Delta H}{H} > \displaystyle \frac{\Delta \Phi_{\bar \nu_e}}{\Phi_{\bar \nu_e}}\;.
\end{equation}

We have roughly estimated that the flux uncertainties of the order of 10 \%,  
that can be achieved at Kamioka with 6 kt$\cdot$yr exposure, 
correspond to 15-30\% in heat uncertainties, consistent with 
the results presented in Ref.~\cite{fiorentini3}.

\subsection{Dependence on the solar mixing angle $\theta_\odot$}
\label{subsec:solar_angle}

So far, we have fixed the solar mixing angle $\theta_\odot$ 
to its current best fitted value.
How our ignorance about the precise value of the solar mixing angle 
can aggravate our results? To answer this question, we perform a fit 
leaving $\sin^2 2 \theta_\odot$ as a free parameter.
For this purpose we will combine present solar neutrino
data with future simulated reactor and geo-antineutrino data.

\begin{figure}
\vglue -2.5cm
\centering\leavevmode
\hglue 2.5cm
\includegraphics[width=14cm]{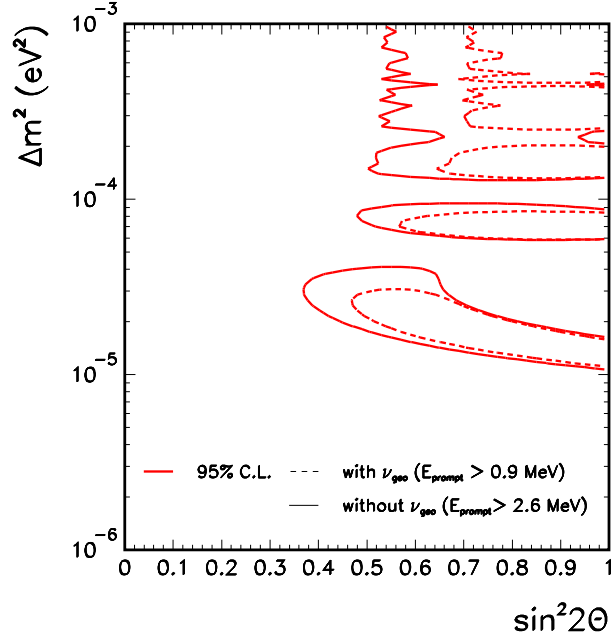}
\vglue -3.8cm
\caption{
Regions in $(\sin^2 2\theta_\odot,\Delta m^2)$ plane allowed by  
KamLAND data alone (without solar neutrino data) for
different thresholds $E_{\text{prompt}} > 0.9$ MeV (including 
geo-antineutrino candidates) 
and $E_{\text{prompt}} > 2.6$ MeV 
(without geo-antineutrinos contribution). We have imposed 
$M(\text{Th})/M(\text{U})=3.8$ in the fit.
} 
\label{kamgeo_sin}
\end{figure}

As a first step, 
to see how the observation of geo-antineutrinos can affect the 
determination of the mixing angle, we have analyzed the present KamLAND 
reactor antineutrino data (17 bins) allowing for geo-antineutrino  
contributions in the 
first 4 bins, as performed by the authors of Ref.~\cite{fiorentini2} leaving 
Th and U antineutrino fluxes free but imposing that the ratio between the 
total U and Th mass in the Earth is fixed to be constant, 
$M(\text{Th})/M(\text{U})=3.8$, as in the end of the previous section.
In Fig.~\ref{kamgeo_sin}, we show the result we have obtained. 
In agreement with the result in Ref.~\cite{fiorentini2}, the 
allowed region becomes somewhat smaller when we include
events which can be interpreted as geo-antineutrinos.
However, we have confirmed that if U and Th contributions 
were treated as independent free parameters, the allowed region 
would not shrink as much, as recently pointed out by Inoue 
in Ref.~\cite{kamland}.  
Some events observed in the energy range $E_{\rm prompt} < 2.6$ MeV 
can be attributed to geo-antineutrinos, but the claim that 
geo-antineutrinos have been observed  can not be made at this point 
due to small statistics (see discussion in previous section).

\begin{figure}
\vglue -3.2cm
\centering\leavevmode
\includegraphics[width=15cm]{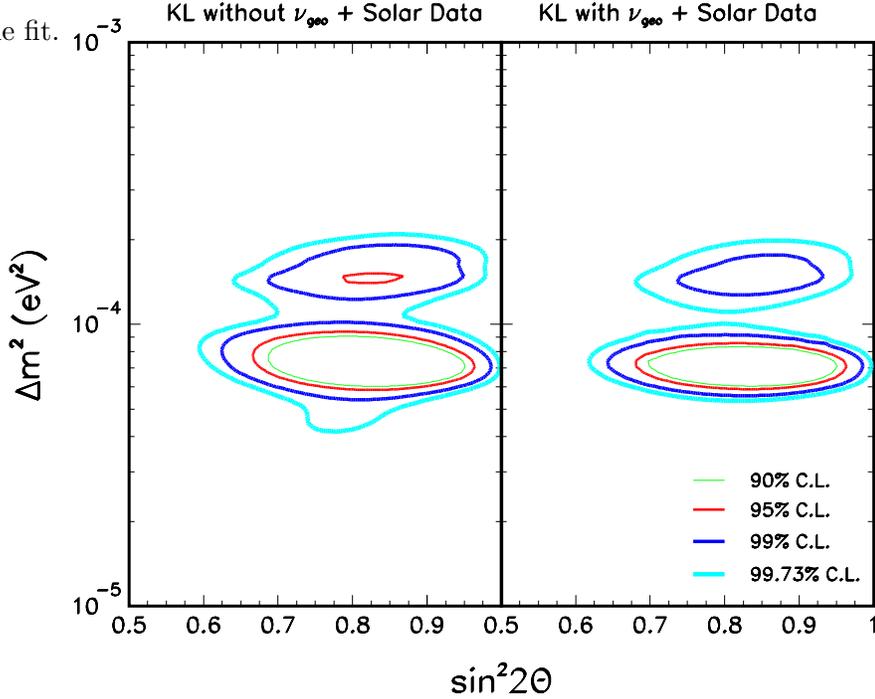}
\vglue -3.8cm
\caption{
Region allowed by all the solar neutrino experiments combined with 
KamLAND data with different thresholds 
$E_{\text{prompt}} > 2.6$ MeV (left panel) and  
$E_{\text{prompt}} > 0.9$ MeV (right panel). 
We have imposed $M(\text{Th})/M(\text{U})=3.8$ in the fit.
} 
\label{totgeo}
\end{figure}

We further proceed, by combining the KamLAND data with the current solar
neutrino one. The final result is shown in the right panel of
Fig.~\ref{totgeo}, where we have also presented the result without the
geo-antineutrino constraint in the left panel taken from Ref.~\cite{NTZ03}. 
The allowed region becomes somewhat smaller but essentially has not changed. 
This is quite understandable  as solar neutrino data are dominating in the 
determination of the mixing angle at this point.

Finally, we combine the simulated future reactor and geo-antineutrino 
data with the present solar neutrino data. 
In order to see how the discrimination power of composition models 
depends on our ignorance of the precise value of the solar mixing angle, 
for a given input value (fixed to be the current 
best fit value) as well as $\Phi_{\bar \nu_e}$(U), we perform 
a $\chi^2$ fit leaving $\Phi_{\bar \nu_e}$(U) and $\theta_\odot$ 
as free parameters. In this fit, we assume the ratio between 
the total U and Th mass in the 
Earth to be constant, $M(\text{Th})/M(\text{U})=3.8$, for simplicity.
This is quite reasonable for our purposes here as the discrimination 
power among models does not seem to depend very much on this assumption,  
as we have  discussed in the previous subsection.

\begin{figure}
\centering\leavevmode
\hglue -2.3cm
\includegraphics[width=23cm]{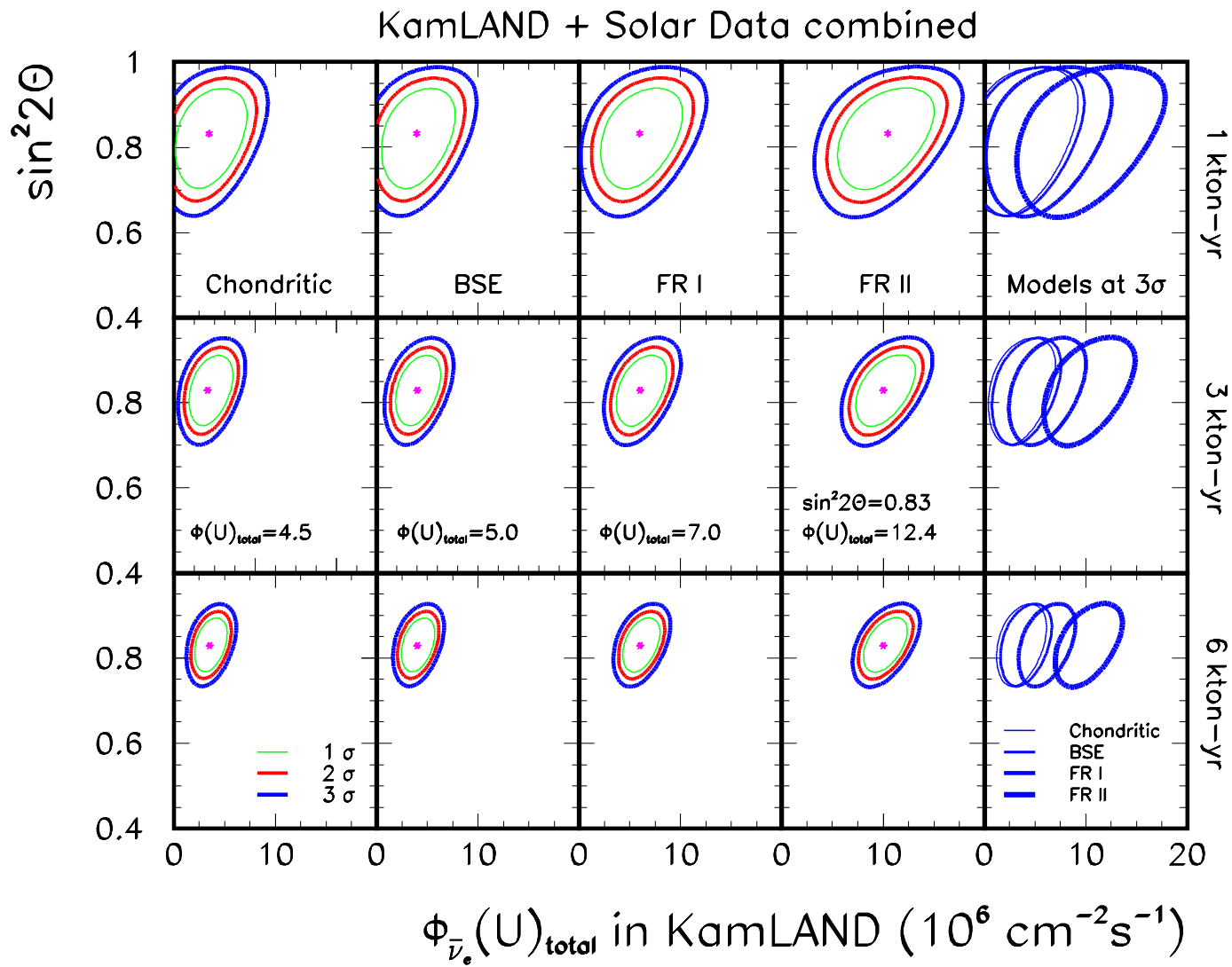}
\vglue -5.5cm
\caption{
 Region allowed at 1, 2 and 3 $\sigma$ in the 
$\Phi_{\bar \nu_e}(\text{U})-\sin^2 2\theta_\odot$ plane, for 4
 different models (in first 4 columns) for 1, 3 and 6 kt$\cdot$ yr 
 of exposures (in different rows). In the 5th column, we have 
 superimposed 3 $\sigma$ allowed regions of 4 different models.  
 Here the flux ratio of
  $\Phi_{\bar \nu_e}(\text{U})$ and $\Phi_{\bar \nu_e}(\text{Th})$ 
 is fixed to be constant.  }
\label{fig:sin_phi_km}
\end{figure}

\begin{figure}
\centering\leavevmode
\hglue -2.3cm
\includegraphics[width=23cm]{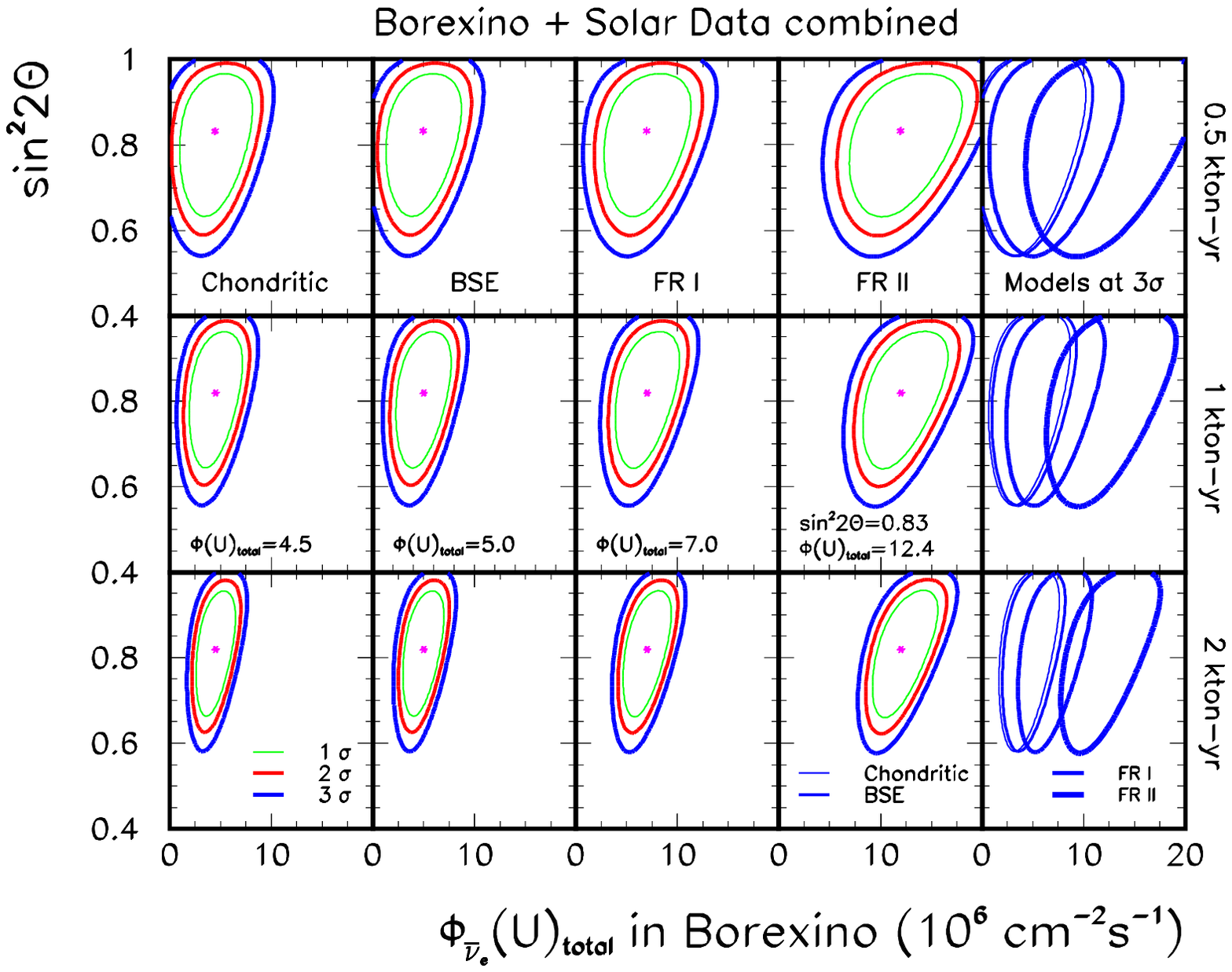}
\vglue -5.5cm
\caption{
Same as in Fig.~\ref{fig:sin_phi_km} but for the Gran Sasso
site. 
}
\label{fig:sin_phi_gs}
\end{figure}

In Figs.~\ref{fig:sin_phi_km} and \ref{fig:sin_phi_gs} we plot the 
allowed region for $\Phi_{\bar \nu_e}$(U) and $\sin^2 2\theta_\odot$
for the Kamioka and Gran Sasso sites.
The reduction of the allowed range of $\sin^2 2 \theta_\odot$ from 
1 to 6 kt$\cdot$yr exposure is essentially due to reactor antineutrinos. 
We observe that our ignorance on the exact value of $\theta_\odot$ does not 
influence very much the determination of $\Phi_{\bar \nu_e}$(U), so 
our conclusions in the previous subsection do not essentially change.

\section{Discussion and Conclusion}
\label{sec:conclusion}
The amount of radioactive elements in the Earth is not well known. The total 
quantity of U and Th in the Earth, however, can be directly measured by  
neutrino detectors.  Presently  KamLAND data imply that the Earth 
radiogenic heat output can be anything between 0 and 110 TW~\cite{kamland}
and all Earth composition models are compatible~\cite{fiorentini2}. 
We have studied how this can be improved by future data.

We have investigated to which extent U and Th 
geo-antineutrino fluxes can be determined by neutrino detectors in a decade of 
exposure. We have considered  KamLAND at Kamioka and 
Borexino at Gran Sasso as our reference detectors and sites. 
We have found, as we showed in Table I, that within a few years with a
relatively small amount of exposure, it is possible to establish the
presence of geo-antineutrinos unless their flux is significantly
smaller than expected.  
However, to discriminate among different Earth composition models, 
considerably longer exposure is required. We found
that in 6 kt$\cdot$yr at Kamioka, the total U flux can be determined
within less than 10~\% uncertainty, independently of 
the model we considered.  
The precision it can be determined at Gran Sasso, after 2 kt$\cdot$yr
exposure, is slightly larger than 10~\%. 
We note that at Gran Sasso the same sensitivity to geo-antineutrino 
flux determination can be achieved with substantially smaller 
exposure due to much lower reactor antineutrino background.

We observe that our ignorance on the exact value of $\theta_\odot$
does not influence very much the determination of the geo-antineutrino
flux. However, it is very important to know the local variation of the
Earth crustal thickness as well as concentration of U and Th in the
region close to the detector with better than $2^\circ \times 2^\circ$
resolution, to be able to accurately translate the measured flux into
amounts of U and Th.

The determination of the radiogenic component of the Earth heat generation 
is of great geophysical interest. Experiments such as KamLAND and Borexino 
can open a new window to survey the internal structure and dynamics 
of our planet, leading to the birth of neutrino geophysics.

\begin{acknowledgments}
This work was supported by Funda{\c c}{\~a}o de Amparo
{\`a} Pesquisa do Estado de S{\~a}o Paulo (FAPESP) and Conselho
Nacional de  Ci{\^e}ncia e Tecnologia (CNPq).
\end{acknowledgments}



\newpage 

\newpage

\end{document}